\documentclass[manuscript,sigconf]{acmart}

\AtBeginDocument{%
  \providecommand\BibTeX{{%
    \normalfont B\kern-0.5em{\scshape i\kern-0.25em b}\kern-0.8em\TeX}}}

\setcopyright{acmlicensed} 
\copyrightyear{2025}
\acmYear{2025}
\acmConference[CHI '25]{CHI Conference on Human Factors in Computing Systems}{April 26-May 1, 2025}{Yokohama, Japan}
\acmBooktitle{CHI Conference on Human Factors in Computing Systems (CHI '25), April 26-May 1, 2025, Yokohama, Japan}

\acmISBN{}
\acmDOI{} 
\acmPrice{} 

\usepackage{algorithm}
\usepackage{algpseudocode}
\usepackage{graphicx}
\usepackage{booktabs}
\usepackage{array}
\usepackage{caption}
\usepackage{multirow}  
\usepackage{float}

\begin{document}


\title{Toward AI-driven Multimodal Interfaces for Industrial CAD Modeling}

\thanks{\textit{Presented at the Bridging HCI and Industrial Manufacturing Workshop, CHI Conference on Human Factors in Computing Systems (CHI '25), April 26-May 1, 2025, Yokohama, Japan.}}

\author{Jiin Choi}
\authornote{These authors contributed equally to this work.}
\affiliation{%
  \institution{Design Informatics Lab, Hanyang University}
  \streetaddress{222, Wangsimni-ro, Seongdong-gu}
  \city{Seoul}
  \country{Republic of Korea}}
\postcode{04763}
\email{jiin4900@gmail.com}

\author{Yugyeong Jang}
\authornotemark[1] 
\affiliation{%
  \institution{Design Informatics Lab, Hanyang University}
  \streetaddress{222, Wangsimni-ro, Seongdong-gu}
  \city{Seoul}
  \country{Republic of Korea}}
\postcode{04763}
\email{jjang020828@gmail.com}

\author{Kyung Hoon Hyun}
\authornote{Corresponding author.}
\affiliation{%
  \institution{Design Informatics Lab, Hanyang University}
  \streetaddress{222, Wangsimni-ro, Seongdong-gu}
  \city{Seoul}
  \country{Republic of Korea}}
\email{hoonhello@gmail.com}

\authorsaddresses{}
\settopmatter{printacmref=false} 
\begin{abstract}
AI-driven multimodal interfaces have the potential to revolutionize industrial 3D CAD modeling by improving workflow efficiency and user experience. However, the integration of these technologies remains challenging due to software constraints, user adoption barriers, and limitations in AI model adaptability. This paper explores the role of multimodal AI in CAD environments, examining its current applications, key challenges, and future research directions. We analyze Bayesian workflow inference, multimodal input strategies, and collaborative AI-driven interfaces to identify areas where AI can enhance CAD design processes while addressing usability concerns in industrial manufacturing settings.

\end{abstract}

\newcommand{\longbar}{
    \tikz[baseline] \draw[thick] (0,0) -- (0,2.5ex);%
}




\maketitle

\section{Introduction}
The industrial design and manufacturing sectors increasingly rely on 3D CAD (Computer-Aided Design) systems to create complex, high-precision models. As these systems evolve, integrating AI-driven automation and multimodal interface, such as voice commands, gesture recognition, and sketch-based interaction, has emerged as a promising direction to enhance efficiency and usability\cite{Niu_2022,kaye2024hci}. However, challenges related to workflow complexity, user adoption, and AI adaptability remain significant barriers to widespread implementation.

Traditional CAD software relies on structured command-based interactions, requiring extensive expertise and manual effort \cite{lukavcevic2023eeg}. The introduction of AI-driven multimodal input methods has the potential to streamline repetitive modeling tasks, optimize design processes, and improve collaboration within industrial teams. Despite these advantages, the transition from conventional CAD workflows to AI-enhanced interfaces introduces new usability concerns, including adaptability to diverse user expertise levels and seamless integration with existing industry-standard tools such as Rhino and SolidWorks.

This paper explores the current applications, challenges, and future directions of AI-driven multimodal interfaces in industrial 3D CAD systems. We analyze how Bayesian workflow inference, adaptive UX strategies, and collaborative AI-powered modeling tools can contribute to more efficient and intuitive CAD environments. Furthermore, we discuss the implications of these technologies for industrial design practices and propose research directions to address existing barriers to adoption.

\section{Multimodal AI in Industrial CAD: Current Applications
}
The integration of AI-driven multimodal interfaces in industrial CAD systems is rapidly evolving, offering new ways to enhance user interaction, streamline workflows, and improve design efficiency (Table \ref{tab:ai_cad_multimodal}). Below, we outline the key applications and emerging trends in this domain.

\begin{table*}[h]
    \centering
    \setlength{\tabcolsep}{6pt} 
    \begin{tabular}{m{4cm} m{4.5cm} m{6.5cm}} 
        \toprule
        \textbf{Category} & \textbf{AI-Driven Application} & \textbf{Description} \\
        \midrule
        \multirow{3}{4cm}{\raggedright \textbf{AI-enhanced CAD Automation} \\ 
        AI-driven automation in CAD enhances efficiency by predicting commands, dynamically adjusting parameters, and enabling generative design exploration.} & 
        Bayesian Command Prediction & 
        AI models analyze past user interactions to predict and recommend the next logical CAD operations~\cite{jang2024advancing}. \\
        \cmidrule(lr){2-3}
        & Automated Parametric Adjustments & 
        AI-driven constraints and optimization techniques refine geometric parameters dynamically~\cite{Jones_2023}. \\
        \cmidrule(lr){2-3}
        & AI-guided Design Exploration & 
        Generative AI enables users to explore alternative design configurations based on predefined constraints and objectives~\cite{Buonamici_2020, Li_2024, choi2025genpara}. \\
        \midrule
        \multirow{3}{4cm}{\raggedright \textbf{Multimodal Interaction in 3D Design} \\ 
        Multimodal AI allows designers to interact with CAD software using alternative input methods like voice, gestures, and sketches, improving usability and accessibility.} & 
        Voice-based Modeling Commands & 
        Natural language processing (NLP) enables users to create and modify models through spoken instructions~\cite{Nanjundaswamy_2013, hucapp21, Plumed_2020}. \\
        \cmidrule(lr){2-3}
        & Text-to-CAD Generation & 
        Large language models (LLMs) enable users to generate CAD models from textual descriptions, allowing automated model creation and reverse engineering of design data~\cite{du2024text2bim, rukhovich2024cad}. \\
        \cmidrule(lr){2-3}
        & Gesture and Motion Controls & 
        Motion-tracking technologies allow users to manipulate 3D objects in immersive design environments~\cite{Huang_2018, Vuletic_2021}. \\
        \cmidrule(lr){2-3}
        & Sketch-to-3D Conversion & 
        AI-powered recognition tools transform freehand sketches into structured CAD models, bridging the gap between ideation and digital modeling~\cite{Li_2020, Li_2022, you2024img2cad, chen2024img2cad, lee2024impact, li2025meshpad}. \\
        \midrule
        \multirow{3}{4cm}{\raggedright \textbf{AI-driven Collaboration in CAD} \\ 
        AI-powered tools enhance collaboration in industrial CAD by improving co-design processes, detecting errors in real time, and facilitating cloud-based teamwork.} & 
        AI-assisted Co-Design Systems & 
        AI supports multi-user environments by predicting modifications and ensuring design consistency~\cite{Tang_2024}. \\
        \cmidrule(lr){2-3}
        & Real-time AI-driven Error Detection & 
        Automated systems identify and correct modeling inconsistencies to maintain design integrity~\cite{Mandorli_2021, Huang_2015}. \\
        \cmidrule(lr){2-3}
        & Cloud-based AI Integration & 
        Web-based platforms enable remote collaboration, leveraging AI for version control and optimization~\cite{Robertson_2022}. \\
        \bottomrule
    \end{tabular}
    \caption{AI-driven Multimodal Interaction in CAD: Key Topics and Descriptions}
    \label{tab:ai_cad_multimodal}
\end{table*}
\section{Key Challenges in AI-driven Multimodal CAD Interfaces}

\subsection{Usability and User Adoption}
Despite the potential benefits of multimodal AI in CAD environments, user resistance remains a significant barrier. Designers and engineers are accustomed to traditional keyboard-mouse interfaces and may be reluctant to adopt voice, gesture, or sketch-based inputs\cite{Bhuiyan_2011}. Furthermore, AI-driven automation may introduce a steep learning curve, requiring users to develop new interaction paradigms. For AI adoption to succeed, it is critical to design interfaces that are intuitive, adaptable to different expertise levels, and provide clear explanations for AI-generated recommendations.

\subsection{Model Adaptability and Data Constraints}
AI models for CAD must generalize across diverse design environments, from automotive engineering to architectural modeling. However, industrial datasets are often proprietary, making it difficult to train robust AI systems on real-world CAD data. Additionally, AI models must be capable of handling variations in user preferences, modeling styles, and domain-specific constraints\cite{zou2024intelligent}. Addressing these challenges requires developing adaptive AI frameworks that can learn from limited data while maintaining high levels of accuracy and reliability.

\subsection{Collaboration and Industry Integration}
Seamless integration with existing CAD software ecosystems such as Rhino, SolidWorks, and AutoCAD is essential. However, interoperability remains a challenge due to differences in file formats, modeling approaches, and scripting languages across platforms. Addressing this requires standardized APIs, cross-platform integration strategies, and collaboration among software developers, researchers, and industry professionals. Furthermore, to fully realize the potential of multimodal CAD interfaces, it is crucial to address key challenges such as workflow complexity, usability, and AI model adaptability. Effective industry integration will be a key factor in driving practical adoption and innovation in AI-assisted CAD environments.

\section{Enhancing Industrial CAD with AI-Driven Multimodal Interaction }
\subsection{Data-Driven AI Optimization for CAD Modeling}
Industrial CAD systems demand precision and efficiency, but AI adoption faces challenges such as data scarcity, proprietary constraints, and industry variability. To overcome these, data-driven optimization techniques are essential. Transfer learning accelerates AI adaptation by leveraging existing design repositories, while synthetic data generation broadens AI applicability across diverse industrial contexts.

Beyond data augmentation, workflow inference enhances CAD efficiency. Bayesian inference refines AI predictions through continuous updates, reducing errors and improving design automation. AI-assisted analysis further streamlines workflows by identifying inefficiencies, optimizing parameters, and refining iterative processes. Integrating these techniques can enhance CAD automation, accelerate production cycles, and reduce costs.

\subsection{Multimodal Interaction for Adaptive CAD UX}
Traditional CAD interfaces rely on keyboard and mouse inputs, but multimodal interaction offers more intuitive and adaptive alternatives. A key challenge is the variation in user expertise—while experienced designers prefer command-driven workflows, novice users benefit from visual and natural input methods. Adaptive UX strategies help bridge this gap by dynamically adjusting interfaces based on user proficiency, project complexity, and design intent.

Multimodal interfaces leverage voice, gestures, and sketch-based inputs to enhance CAD usability. For example, voice-based commands allow users to describe modeling operations verbally, gesture recognition enables 3D object manipulation, and sketch-to-3D conversion simplifies the transition from conceptual design to structured modeling \cite{lee2024impact}. AI systems can analyze user behavior in real-time and automatically adjust the interface, ensuring seamless interactions across different experience levels. By integrating multimodal interaction, CAD software becomes more adaptive, intuitive, and efficient, reducing the learning curve and improving workflow accessibility.

\subsection{AI-CAD System Integration and Industry Collaboration}
For AI-driven multimodal interfaces to be widely adopted in industrial CAD, seamless integration with platforms like Rhino, SolidWorks, and AutoCAD is essential. Standardized APIs, plug-in architectures, and interoperability frameworks are critical for ensuring compatibility across diverse software ecosystems. Additionally, real-time synchronization between AI-driven tools and traditional CAD workflows can enhance adoption and minimize disruptions.

A key challenge in AI-driven CAD adoption is bridging the gap between academia and industry. Research should focus on developing collaborative frameworks for knowledge transfer, including industry-specific AI benchmarks, anonymized CAD datasets, and real-world deployment strategies. Addressing concerns such as data security, intellectual property protection, and compliance with design standards will further accelerate adoption.

By enhancing AI-CAD integration and fostering industry-academic collaboration, AI-driven multimodal interfaces can transition from theoretical concepts to practical tools that improve efficiency, streamline workflows, and drive innovation in industrial design and manufacturing.

\section{Conclusion}
Multimodal AI is reshaping industrial CAD by enhancing design automation, improving user interaction, and optimizing workflows. Bayesian inference, adaptive UX strategies, and collaborative AI tools contribute to more intuitive and efficient design environments.
However, challenges remain, including workflow integration, user adoption, and the scalability of AI models across diverse industrial domains. Addressing these issues requires ongoing research into enhancing AI-CAD interoperability, refining multimodal interaction techniques, and developing scalable AI-driven design assistance tools.
Collaboration between academia and industry will be essential in bridging the gap between theoretical advancements and real-world implementation. As AI continues to evolve, its role in CAD automation is expected to expand, enabling more intelligent, adaptive, and efficient design workflows. By overcoming current barriers to adoption and ensuring alignment with industry needs, AI-driven CAD systems have the potential to revolutionize industrial 3D modeling, driving innovation and productivity in manufacturing and design.

\begin{acks}
This work was supported by National Research Foundation of Korea (NRF) grant funded by the Korean government (MSIT; RS-2023-00208542).
\end{acks}

\bibliographystyle{ACM-Reference-Format}
\bibliography{sample-base}

\end{document}